\documentclass[12pt]{article}
\usepackage{amsfonts}
\usepackage{amsmath}
\usepackage{amssymb}
\usepackage{graphicx}
\usepackage{color}
\usepackage[all, knot]{xy}
\usepackage{tikz}
\usepackage{cancel}
\usepackage{ulem}

\usepackage[utf8]{inputenc}
\usepackage{epstopdf}
\usepackage[footnotesize]{caption}
\usepackage{amsthm}
\usepackage{enumitem}
\usepackage{mathrsfs}

\usepackage[margin=3cm]{geometry}

\def \be {\begin{equation}}
\def \ee {\end{equation}}
\def \bea {\begin{eqnarray}}
\def \eea {\end{eqnarray}}
\def \nn {\nonumber}

\def \rr {\raise.35ex\hbox{\small $\prime$}\kern-.17em{\mbox{\large $\imath$}}}

\def \dels {\partial\kern-.6em /\kern.1em}
\def \As {{A\kern-.5em / \kern.5em}}
\def \Ds {D\kern-.7em / \kern.5em}

\def \ks {k\kern-.5em /}
\def \ls {l\kern-.5em /}

\newcommand{\ci}[1]{}

\newcommand{\ba}{\begin{eqnarray}}
\newcommand{\ea}{\end{eqnarray}}
\newcommand{\bal}{\begin{align}}
\newcommand{\eal}{\end{align}}
\newcommand{\bay}[1]{\left(\begin{array}{#1}}
\newcommand{\eay}{\end{array}\right)}

\newcommand{\hide}[1]{}

\newlist{axioms}{enumerate}{2}
\setlist[axioms,1]{label=\textbf{A\arabic{axiomsi}.}, ref=A\arabic{axiomsi}}
\setlist[axioms,2]{label=\textbf{A\arabic{axiomsi}\rlap{\myEnumCounter{axiomsii}}.},%
                   ref=A\arabic{axiomsi}\myEnumCounter{axiomsii},%
                   align=parleft,%
                   leftmargin=0em,%
                   itemsep=1.4ex,%
                   before={\stepcounter{axiomsi}}}
                   
  \usetikzlibrary{decorations.markings}

\begin{document}

\begin{titlepage}
\begin{center}

\textbf{\LARGE
Quantum Entanglement and\\
 Spectral Form Factor 
\vskip.3cm
}
\vskip .5in
{\large
Chen-Te Ma$^{a, b, c, d, e, f}$ \footnote{e-mail address: yefgst@gmail.com} and
Chih-Hung Wu $^{f, g}$ \footnote{e-mail address: chih-hungwu@physics.ucsb.edu}
\\
\vskip 1mm
}
{\sl
$^a$ 
Asia Pacific Center for Theoretical Physics,\\
Pohang University of Science and Technology, 
Pohang 37673, 
Gyeongsangbuk-do, South Korea.\\
$^b$
Guangdong Provincial Key Laboratory of Nuclear Science,\\
 Institute of Quantum Matter,
South China Normal University, Guangzhou 510006, Guangdong, China.
\\
$^c$
 School of Physics and Telecommunication Engineering,\\
 South China Normal University, Guangzhou 510006, Guangdong, China.
\\
$^d$
Guangdong-Hong Kong Joint Laboratory of Quantum Matter,\\
 Southern Nuclear Science Computing Center, 
South China Normal University,\\
 Guangzhou 510006, Guangdong, China.\\
$^e$
The Laboratory for Quantum Gravity and Strings,\\
 Department of Mathematics and Applied Mathematics,
University of Cape Town, Private Bag, Rondebosch 7700, South Africa.
\\
$^f$
Department of Physics and Center for Theoretical Sciences,\\ 
 National Taiwan University, Taipei 10617, Taiwan,
R.O.C.
\\
$^g$
Department of Physics, University of California Santa Barbara, Santa Barbara,\\
 CA 93106-9530, USA.
}\\
\vskip 1mm
\vspace{40pt}
\end{center}
\newpage
\begin{abstract}
We replace a Hamiltonian with a modular Hamiltonian in the spectral form factor and the level spacing distribution function. 
This study establishes a connection between quantities within Quantum Entanglement and Quantum Chaos.
To have a universal study for Quantum Entanglement, we consider the Gaussian random 2-qubit model. 
The maximum violation of Bell's inequality demonstrates a positive correlation with the entanglement entropy. 
Thus, the violation plays an equivalent role as Quantum Entanglement. 
We first provide an analytical estimation of the relation between quantum entanglement quantities and the dip when a subregion only has one qubit.
The time of the first dip is monotone for entanglement entropy. 
The dynamics in a subregion is independent of the initial state at a late time. 
It is one of the signaling conditions for classical chaos. 
We also extend our analysis to the Gaussian random 3-qubit state, and it indicates a similar result.
The simulation shows that the level spacing distribution function approaches GUE at a late time.  
In the end, we develop a technique within QFT to the spectral form factor for its relation to an $n$-sheet manifold.
We apply the technology to a single interval in CFT$_2$ and the spherical entangling surface in $\mathcal{N}=4$ super Yang-Mills theory. 
The result is one for both cases, but the Rényi entropy can depend on the Rényi index. 
For the case of CFT$_2$, it indicates the difference between the continuum and discrete spectrum.
\end{abstract}
\end{titlepage}

\section{Introduction}
\label{sec:1}
\noindent
A mystery of quantum physics is that an absolute prediction of the measurement outcome is generally unobtainable. 
This phenomenon distinguishes a quantum system from its classical counterpart.
An intuitive way to understand this mystery is to introduce the hidden variables in quantum theory.
However, the local hidden variables are incompatible with quantum theory from Bell's inequality in a 2-qubit system \cite{Bell:1964kc,Clauser:1969ny}. 
Retrieving information from a quantum state through its entanglement structure thus becomes the main focus of {\it Quantum Information}. 
\\

\noindent 
The characteristic feature of quantum information theory is {\it Quantum Entanglement}. 
Intuitively, the phenomenon is that each particle cannot be described individually in a quantum state. 
The {\it maximum violation of Bell's inequality} provides a concrete manifestation. 
Then the quantum generalization of Bell's inequality \cite{Cirelson:1980ry,Horodecki:1995} allows a connection to the concurrence of pure state (or entanglement entropy of pure state) \cite{Bennett:1996gf} in an arbitrary 2-qubit system \cite{Verstraete:2001}. 
\\

\noindent 
Many studies in {\it Quantum Entanglement} were only derivable based on numerical approaches. 
Therefore, developing new techniques for obtaining analytical solutions is an urgent task.
Ryu and Takayanagi established the relation between the Anti-de Sitter (AdS) minimum surface and entanglement entropy \cite{Ryu:2006bv}. 
This surface simplifies the calculation of entanglement entropy in CFT.
Although the conformal symmetry has largely constrained possible analytical solutions, the $n$-sheet manifold method employed in the calculation could be uncontrollable \cite{Casini:2010kt,Huang:2014pda}. 
Developing methods beyond or without the replica trick in the {\it AdS/CFT correspondence} \cite{Maldacena:1997re} and  {\it Entanglement in Quantum Field Theory} (QFT) \cite{Calabrese:2004eu} have also received much attention in {\it Quantum Information}.
\\

\noindent
According to a recent observation, the saturation of chaotic bound in the Lyapunov exponent is a necessary condition for
the bulk Einstein gravity theory \cite{Maldacena:2015waa}. 
The maximum chaotic bound also indicates the maximum sensitivity of initial conditions. 
This information in a physical system appears through its interaction with other systems and their dynamics. 
{\it Quantum Entanglement} loses individual dynamics information. 
Hence {\it Quantum Chaos} is a yet probed direction in {\it Quantum Information} for obtaining essential additional information from {\it Quantum Entanglement}.
\\

\noindent
The difficulty in {\it Quantum Chaos} is a closed quantum system because the wavefunction satisfies a linear equation. 
The linear equation cannot offer any irregular information to the dynamics. 
The correspondence principle should imply a classical limit from {\it Quantum Chaos} to {\it Classical Chaos}. 
All information of a quantum system should be in the spectrum.  
One general expectation is that the chaotic spectrum should have random matrix statistics \cite{Dyson:1962es}. 
The spectrum of chaotic theory also indicates level repulsion in the level spacing distribution function \cite{Bohigas:1983er,Bohigas:1983er2}. 
\\

\noindent
Recently, people have replaced the Hamiltonian with a modular Hamiltonian in quantum chaos quantities \cite{DeBoer:2019kdj,Chen:2017yzn}. 
The result shows a similar chaotic bound in the Lyapunov exponent \cite{DeBoer:2019kdj}, and the CFTs with a spherical entangling surface saturate the chaos bound \cite{Casini:2017roe,Huang:2019wzc,Huang:2020cye}. 
This study provides a path from {\it Quantum Entanglement} to {\it Quantum Chaos}. 
\\

\noindent
In this paper, we study the spectral form factor from the opinion of Quantum Entanglement (by replacing the Hamiltonian with a modular Hamiltonian). 
We begin with the Gaussian random 2-qubit model. 
This model has an analytical relation between entanglement entropy and maximum violation of Bell's inequality.
We study the relation between the spectral form factor and the concurrence of a pure state. 
The maximum degree of freedom in a 2-qubit model is just four. 
The random coupling constant generates a statistical ensemble with infinite degrees of freedom \cite{Dyson:1962es,Lau:2018kpa,Lau:2020qnl}. 
Since QFT has a continuum spectrum, it is hard to argue the behavior from random qubit models. 
We develop the technique of calculating the spectral form factor from the $n$-sheet manifold. 
We then use Rényi entropy for a single interval in CFT$_2$ to calculate the spectral form factor as a probe of the entanglement spectrum.
To summarize our results:
\begin{itemize} 
\item The occurring time of the first dip in the spectral form factor is positively correlated to entanglement entropy in the Gaussian random 2-qubit model. 
We also consider the random 3-qubit model and obtain a consistent result.
\item The dynamics is independent of the choice of initial states at the late time. 
It is one of the conditions for classical chaos.
We study the Gaussian random 2- and 3-qubit models. 
The numerical study of the level spacing distribution function shows that the spectrum follows the Gaussian unitary ensemble (GUE).
\item 
When one considers Bell's state, the spectral form factor is one. 
Each eigenvalue of the modular Hamiltonian has the same value in this state.
When we calculate the spectral form factor for a single interval in CFT$_2$, the spectral form factor is also one. 
The result exhibits the difference between the continuum and discrete spectrum because the Rényi entropy can depend on the Rényi index. 
This difference provides a strong constraint to the entanglement spectrum of QFT.
The spectral form factor for the twisted density matrix is also one in $\mathcal{N}=4$ super Yang-Mills theory. 
Here we extend the definition by replacing the reduced density matrix with the twisted one.
Hence the constraint should be helpful for a further study of AdS/CFT correspondence.
\end{itemize}

\noindent
The organization of this paper is as follows: We calculate the spectral form factor and level spacing distribution function in the Gaussian random 2- and 3- qubit models by replacing the Hamiltonian with a modular Hamiltonian in Sec.~\ref{sec:2}. In Sec.~\ref{sec:3}, we then relate the $n$-sheet manifold to calculating the spectral form factor in QFT. 
Finally, we discuss our conclusion in Sec.~\ref{sec:4}.

\section{Modular Hamiltonian}
\label{sec:2}
\noindent
The modular Hamiltonian is defined by
\bea
H_{\mathrm{mod}}\equiv-\ln\rho_A,
\eea
where $\rho_A$ is a reduced density matrix of the region $A$.
We will calculate the spectral form factor and level spacing distribution function by replacing the Hamiltonian with the modular Hamiltonian.
Since we hope to obtain a universal result in Quantum Entanglement, we first consider the simple Gaussian random 2-qubit model. 
The Hamiltonian of the Gaussian random two-qubit model is 
\bea
H_{\mathrm{tq}}\equiv\sum_{j,k}g_{j,k}\sigma_j\otimes\sigma_k,
\eea
The random coupling constant is distributed as $\exp(-g_{j,k}^2/2)/\sqrt{2\pi}$.  
The Pauli matrix is given by:
\bea
\sigma_x^2=\sigma_y^2=\sigma_z^2=I, 
\nn
\eea
\bea
\sigma_x\sigma_y=-\sigma_y\sigma_x, \qquad \sigma_y\sigma_z=-\sigma_z\sigma_y, \qquad \sigma_x\sigma_z=-\sigma_z\sigma_x,
\nn
\eea
\bea
\sigma_x\sigma_y=i\sigma_z, \qquad \sigma_y\sigma_z=i\sigma_x, \qquad \sigma_z\sigma_x=i\sigma_y,
\nn
\eea
\bea
\sigma_x|0\rangle=|1\rangle, \qquad \sigma_x|1\rangle=|0\rangle,
\nn
\eea
\bea
\sigma_y|0\rangle=i|1\rangle, \qquad \sigma_y|1\rangle=-i|0\rangle,
\nn
\eea
\bea
\sigma_z|0\rangle=|0\rangle, \qquad \sigma_z|1\rangle=-|1\rangle,
\eea
where
\bea
|0\rangle\equiv\begin{pmatrix}
1\\
0
\end{pmatrix}; \qquad |1\rangle\equiv\begin{pmatrix}
0\\
1
\end{pmatrix}.
\eea
The initial quantum state is given by
\bea
|\psi\rangle=\sqrt{1-b^2}|00\rangle+b|11\rangle,
\eea
in which $b$ is ranged from 0 to 1. 
We will take the partial trace over the second qubit to obtain the reduced density matrix of region $A$.
\\ 

\noindent 
Since the dimensions of Hilbert space increase, the statistical ensemble should approach the random matrix ensemble. 
We consider the initial state as the generalized GHZ state 
\bea
|\psi\rangle=\sqrt{1-b^2}|000\rangle+b|111\rangle,
\eea
in which $b$ is ranged from 0 to 1. 
We partial trace over the second and third qubits to obtain the reduced density matrix of the region $A$. 
The Hamiltonian of the Gaussian random three-qubit model is 
\bea
H_{\mathrm{tq}}\equiv\sum_{j,k,l}g_{j,k,l}\sigma_j\otimes\sigma_k\otimes\sigma_l,
\eea
The random coupling constant is distributed as $\exp(-g_{j,k,l}^2/2)/\sqrt{2\pi}$. 
The 3-qubit result indicates GUE. 

\subsection{Spectral Form Factor} 
\noindent
We first perform some analytical studies for the spectral form factor. 
The spectral form factor is \cite{Chen:2017yzn}
\bea
g_A(\tau)=\frac{1}{N_A^2}\sum_{j, k=1}^{N_A}e^{-i\tau(\lambda_j-\lambda_k)},
\eea
where $N_A$ is the number of eigenvalues of a modular Hamiltonian.
The $g_A$ is one if all eigenvalues are the same after applying a time evolution. 
 The $\lambda_j(t)$ is the eigenvalue of a modular Hamiltonian of the region $A$, and $\tau$ is the Fourier time. 
The dynamics of the spectral form factor demonstrates by the time $t$ dependence of eigenvalues. 
 The Bell's state is for the choice
 \bea
 b=\frac{\sqrt{2}}{2},
 \eea
which refers to a pure state on two qubits. 
This state has maximum entanglement entropy. 
 \\
 
 \noindent
 When a subregion $A$ only has one qubit, the eigenvalues of a reduced density matrix $\tilde{\lambda}_1$ and $\tilde{\lambda}_2$ would satisfy the following condition \cite{Verstraete:2001}:
 \bea
 \tilde{\lambda}_1+\tilde{\lambda}_2=1; \qquad \tilde{\lambda}_1^2+\tilde{\lambda}_2^2=\mathrm{Tr}_A\rho_A^2. 
 \eea
 When we choose 
 \bea
 \tilde{\lambda}_1>\tilde{\lambda}_2,
 \eea
  the eigenvalues are given by \cite{Verstraete:2001}:
 \bea
 \tilde{\lambda}_1&=&\frac{1+\sqrt{2\mathrm{Tr}_A(\rho_A^2)-1}}{2};
 \nn\\
   \tilde{\lambda}_2&=&\frac{1-\sqrt{2\mathrm{Tr}_A(\rho_A^2)-1}}{2}.
 \eea
 Hence the spectral form factor for the initial state is: 
 \bea
 g_A(\tau)&=&\frac{1}{2}+\frac{1}{2}\cos\bigg(\tau\ln\frac{\tilde{\lambda}_1}{\tilde{\lambda}_2}\bigg)
 \nn\\
&=&\frac{1}{2}+\frac{1}{2}\cos\bigg(\tau\ln\frac{1+\sqrt{2\mathrm{Tr}_A(\rho_A^2)-1}}{1-\sqrt{2\mathrm{Tr}_A(\rho_A^2)-1}}\bigg).
 \eea 
 Hence we observe that the spectral form factor is in the form of an oscillating function.  
By including a time evolution, we would have to consider the effect of the random coupling constant. 
It would give an averaging behavior for all oscillating functions from different random configurations. 
 \\
 
 \noindent 
The minimum point of $g_A$ occurs at
\bea
\tau_{\mathrm{min}}=\frac{n\pi}{G}, 
\eea
where
\bea
G\equiv\ln\frac{\tilde{\lambda}_1}{\tilde{\lambda}_2}.   
\eea 
The $n$ is an integer. 
We choose $n=1$ to obtain the occurring time of the first dip in the spectral form factor. 
When $G$ increases, $\mathrm{Tr}_A\rho_A^2$ also increases. 
This result implies that Quantum Entanglement generates the occurring time of the first dip.  
For the Bell's state, the first dip should be at infinity. 
Hence the spectral form factor is one. 
Now we have established a rough interpretation of the behavior of the spectral form factor from Quantum Entanglement.
\\

\noindent
Now we discuss the relation of the concurrence of a pure state \cite{Bennett:1996gf}
\bea
C_A\equiv\sqrt{2(1-\mathrm{Tr}_A\rho_A^2)}
\eea
and Bell's inequality in the Gaussian random 2-qubit model.
When $G$ decreases, the concurrence of a pure state increases (which also implies that entanglement entropy increases). 
It is interesting to note that the maximum violation of Bell's inequality \cite{Verstraete:2001}
\bea
\gamma\equiv\mathrm{max}_{\mathcal{B}}\mathrm{Tr}(\rho\mathcal{B})\le 2\sqrt{1+C_A^2},
\eea
where
\bea
\mathcal{B}\equiv 
\big(\vec{b}\cdot\vec{\sigma}\big)\otimes \bigg(\big(\vec{a}+\vec{\tilde{a}}\big)\cdot\vec{\sigma}\bigg)
+
\big(\vec{\tilde{b}}\cdot\vec{\sigma}\big)\otimes \bigg(\big(\vec{a}-\vec{\tilde{a}}\big)\cdot\vec{\sigma}\bigg),
\nn\\
\eea
 increases with the concurrence of a pure state. 
 The $\vec{a}$, $\vec{\tilde{a}}$, $\vec{b}$, and $\vec{\tilde{b}}$ are unit vectors, and $\vec{\sigma}\equiv(\sigma_x, \sigma_y, \sigma_z)$ is a vector of the Pauli matrix. 
 Hence a 2-qubit system is qualitatively consistent with Quantum Entanglement. 
 \\
 
\begin{figure*}
\includegraphics[width=1.\textwidth]{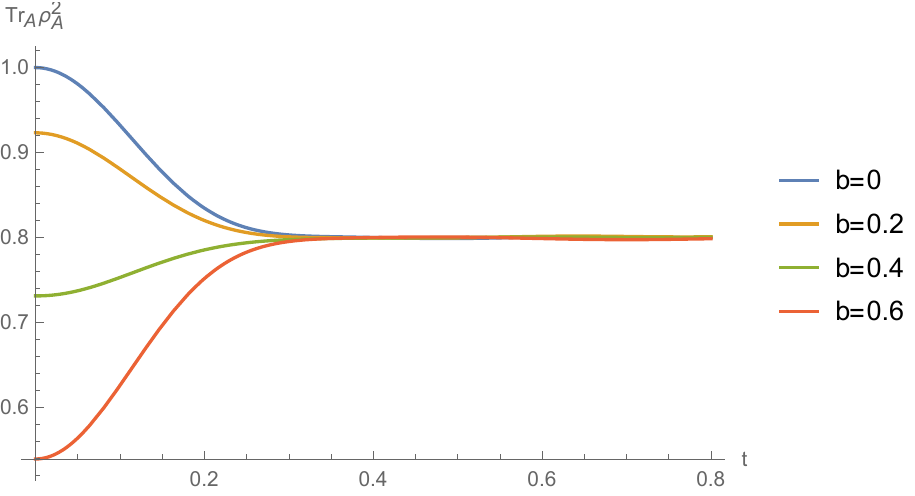}
\caption{We observe that $\mathrm{Tr}_A\rho_A^2$ is monotonically increasing or decreasing with time and then converges to a stable final value.
\label{Renyi2.pdf}}
\end{figure*}

\begin{figure*}
\includegraphics[width=1.\textwidth]{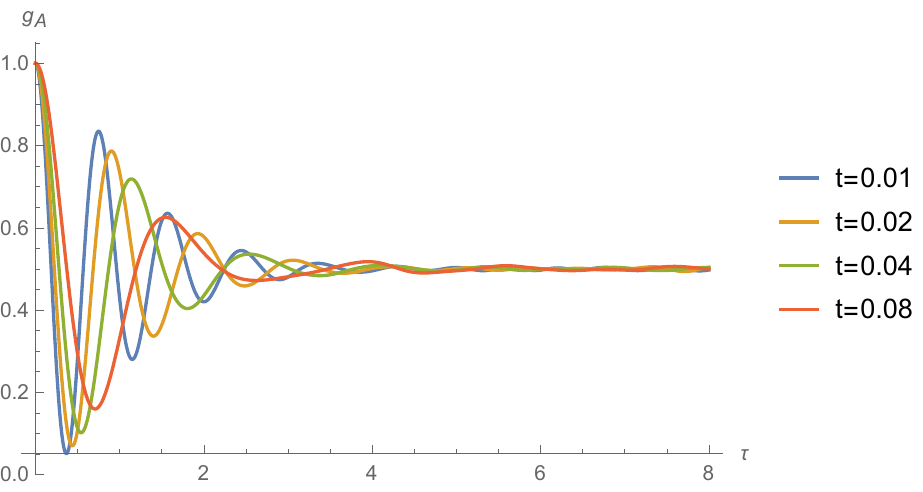}
\includegraphics[width=1.\textwidth]{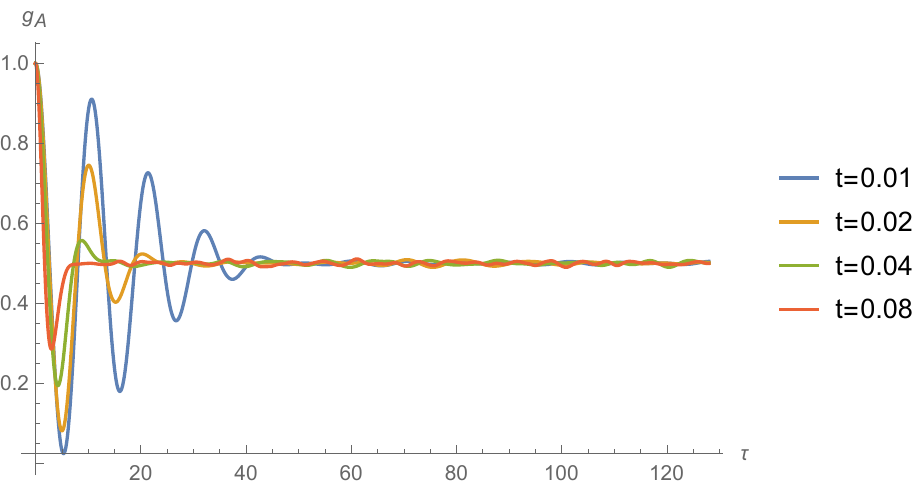}
\caption{The upper figure is for $b=0$, and the lower figure is for $b=0.6$.
We show that the occurring time of the first dip happens earlier when the value of $\mathrm{Tr}_A\rho_A^2$ is smaller.
\label{b.pdf}}
\end{figure*}

\begin{figure*}
\includegraphics[width=1.\textwidth]{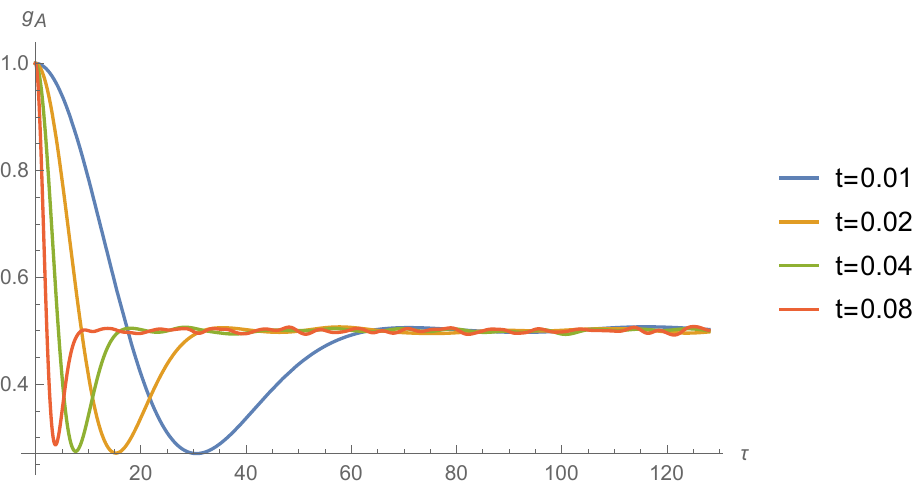}
\caption{The initial state is the Bell's state with $b=\sqrt{2}/2$.  
We show that the occurring time of the first dip happens earlier when the value of $\mathrm{Tr}_A\rho_A^2$ is smaller.
\label{GHZ.pdf}}
\end{figure*}

\noindent
We first calculate $\mathrm{Tr}_A\rho_A^2$ in Fig. \ref{Renyi2.pdf}. 
When its value becomes large, it implies that the case has more configurations with a larger $\mathrm{Tr}_A\rho_A^2$. 
The naive estimation should lead to an early occurring time of the first dip. 
The numerical study confirms this estimation as in Figs. \ref{b.pdf} and \ref{GHZ.pdf}. 
We consider $2^{13}$ random configurations for all simulations in this paper.
\\

\begin{figure*}
\includegraphics[width=1.\textwidth]{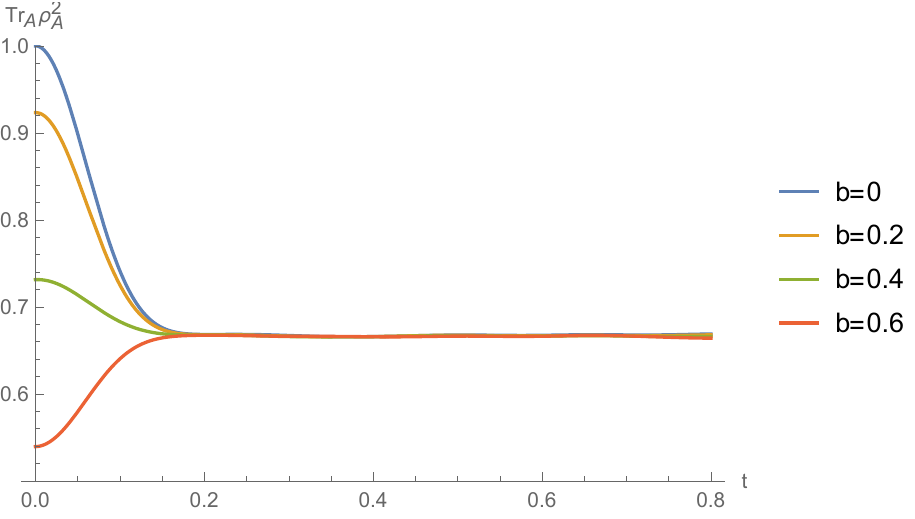}
\caption{We observe that $\mathrm{Tr}_A\rho_A^2$ is monotonically increasing or decreasing with time and then converges to a stable final value.
\label{Renyi2-3qubit.pdf}}
\end{figure*}

\begin{figure*}
\includegraphics[width=1.\textwidth]{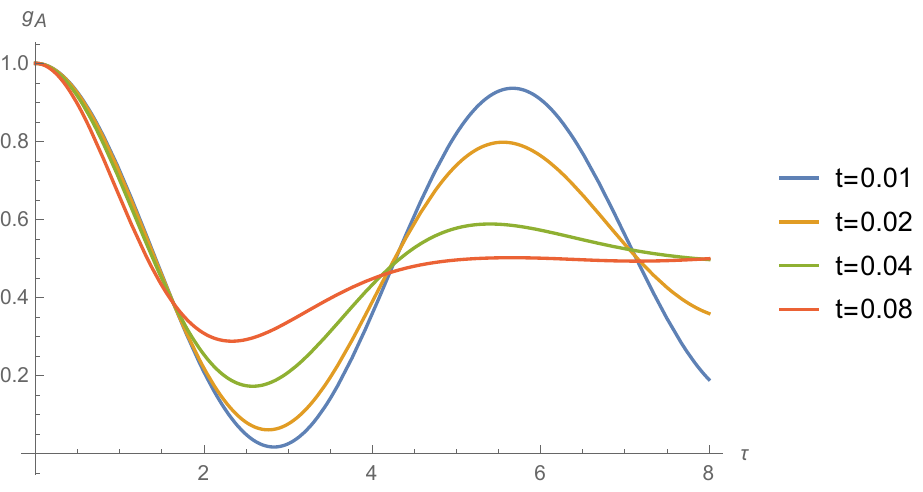}
\includegraphics[width=1.\textwidth]{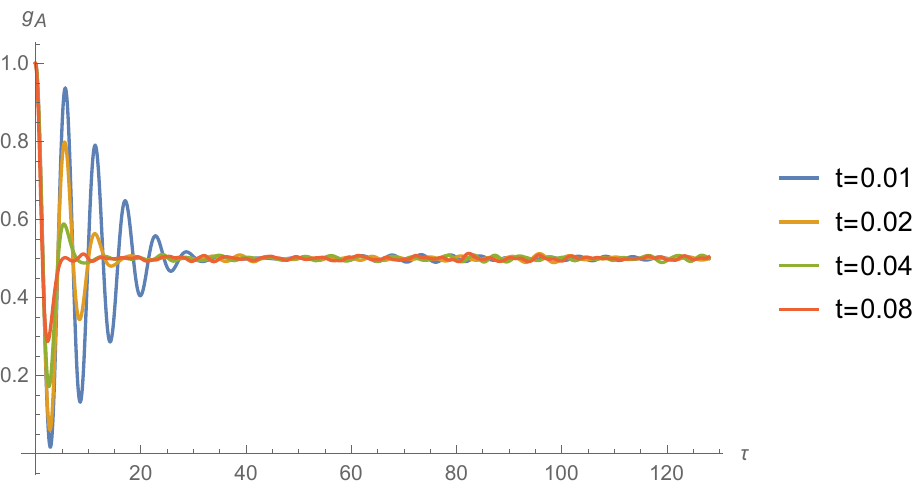}
\caption{The upper figure is for $b=0$, and the lower figure is for $b=0.6$.
We show that the occurring time of the first dip happens earlier when the value of $\mathrm{Tr}_A\rho_A^2$ is smaller.
\label{b-3qubit.pdf}}
\end{figure*}

\begin{figure*}
\includegraphics[width=1.\textwidth]{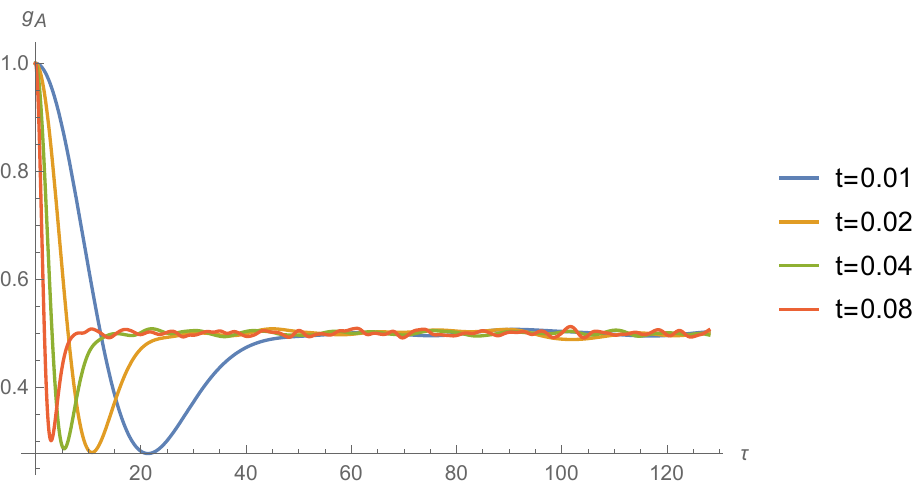}
\caption{The initial state is the generalized GHZ state with $b=\sqrt{2}/2$.  
We show that the occurring time of the first dip happens earlier when the value of $\mathrm{Tr}_A\rho_A^2$ is smaller.
\label{GHZ-qubit.pdf}}
\end{figure*}

\noindent
Now we plot the 3-qubit result in Figs. \ref{Renyi2-3qubit.pdf}, \ref{b-3qubit.pdf}, and \ref{GHZ-qubit.pdf}. 
We get a similar result to the two-qubit case. 
In Fig. \ref{Renyi2-3qubit.pdf}, the Rényi-2 entropy approaches the equilibrium state earlier than the two-qubit random model. 
We expect that it is one of the signals for approaching chaotic behavior. 
Later we will show the level spacing distribution and find GUE. 
This result should be clear for our argument. 

\subsection{Level Spacing Distribution Function}
\noindent
From Figs. \ref{Renyi2.pdf} and \ref{Renyi2-3qubit.pdf}, we observe that the late-time behavior is independent of the choice of initial conditions. 
The observation implies that the late-time dynamics is also independent of the choice of the initial conditions. 
Such independence is one of the conditions in classical chaos.
\\

\noindent
 A natural question to ask is whether the 2-qubit system is too simple. 
 In an oversimplified system, Quantum Entanglement is possibly the only factor for chaos.
 The chaotic features may only be relevant to its entanglement properties in a simplified system. 
 To further examine this question, we should study the late-time behavior of the level spacing distribution function and compare it with the random matrix theory \cite{Dyson:1962es}.
 \\

\begin{figure*}
\includegraphics[width=0.5\textwidth]{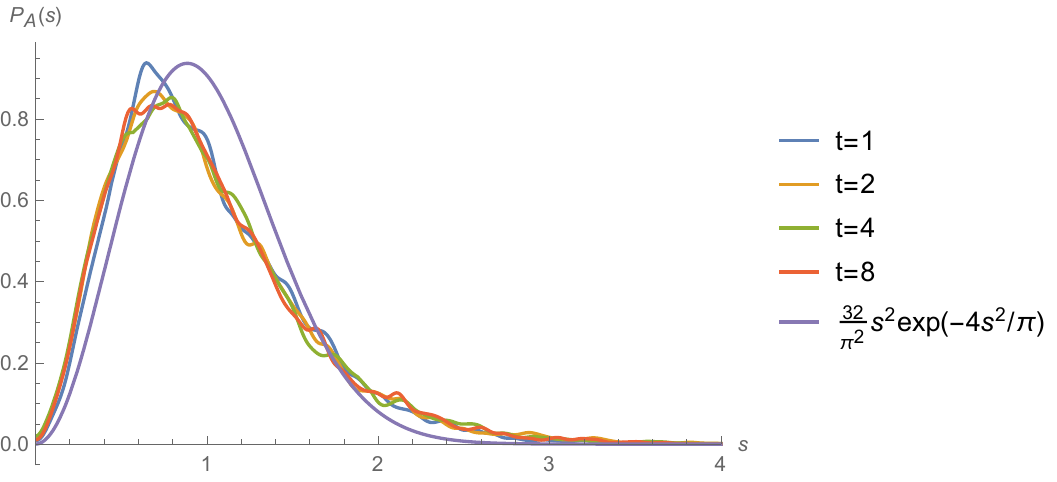}
\includegraphics[width=0.5\textwidth]{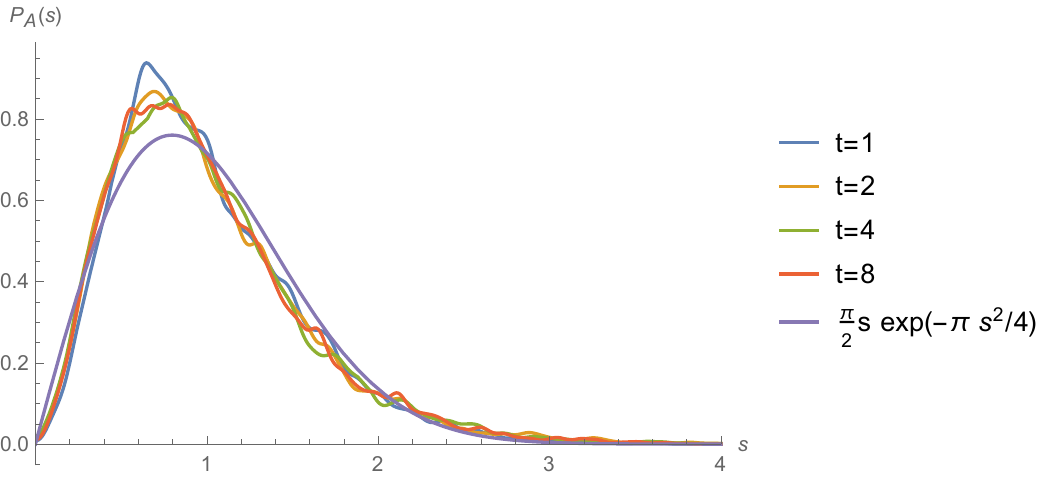}
\caption{The left figure is for the comparison with the Gaussian Unitary Ensemble (GUE), while the right figure is for the Gaussian Orthogonal Ensemble (GOE). The initial state is Bell's state.
\label{G.pdf}}
\end{figure*}

\begin{figure*}
\includegraphics[width=0.5\textwidth]{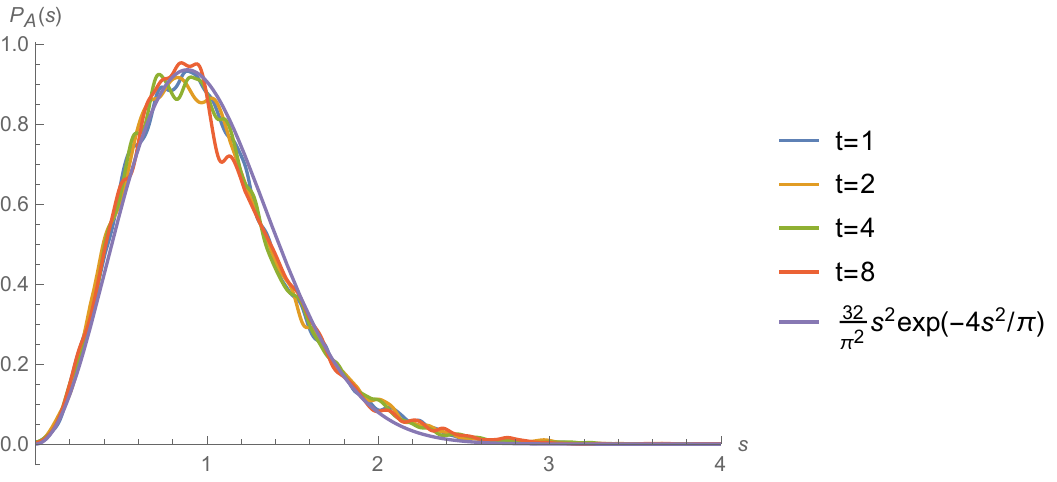}
\includegraphics[width=0.5\textwidth]{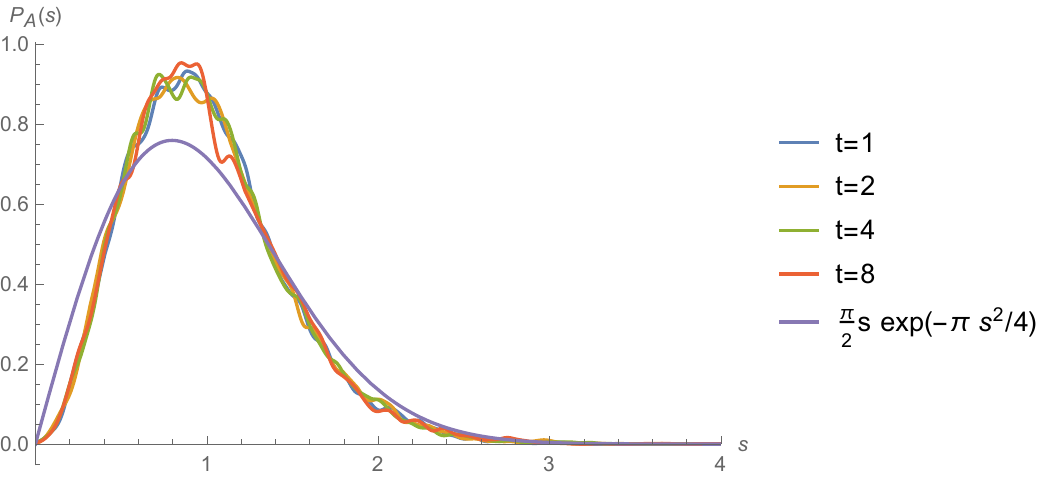}
\caption{The left figure is for the comparison with the Gaussian Unitary Ensemble (GUE), while the right figure is for the Gaussian Orthogonal Ensemble (GOE). The initial state is the generalized GHZ state.
\label{G-3qubit.pdf}}
\end{figure*}

\noindent
Note that the level spacing distribution function is a probability distribution of the normalized energy difference
 \bea
 s\equiv\frac{S}{\langle S\rangle},
 \eea
 where $S$ is the energy difference for each random configuration, and 
 $\langle S\rangle$ is a random average of the energy difference. 
 We have used Bell's state as an initial state. 
 From Fig. \ref{G.pdf}, the system does not have random matrix statistics.
The expectation is that no random matrix statistics should occur due to too few qubits in the model. 
Now we show the 3-qubit result in Fig. \ref{G-3qubit.pdf}. 
Hence we expect that the random coupling constant generates a statistical ensemble. 
Because Quantum Entanglement has no classical analogy. 
By "Wigner's surmise", we expect that the behavior is due to the complexity. 
The random coupling generates the required complexity for a system. 
We should observe the result of the level spacing function approaching the random matrix theory when a system has enough complexity \cite{Dyson:1962es}. 
This conclusion should not be affected by introducing more qubits \cite{Lau:2018kpa}.

\section{QFT}
\label{sec:3}
\noindent 
We observe the universal result in random qubit models. 
However, it is hard to expect the extension to QFT due to the spectrum being a continuum. 
Hence we need to develop the computational technique for the purpose of studying QFT.  
\\

\noindent
Now we demonstrate how the analytical continuation helps the calculation of the spectral form factor with a modular Hamiltonian. 
We calculate the spectral form factor with a modular Hamiltonian for a single interval with length $L$ in CFT$_2$:
\bea
g_A(\tau)&=&\frac{1}{N_A^2}\sum_{j, k=1}^{N_A}e^{-i\tau(\lambda_j-\lambda_k)}
\nn\\
&=&\frac{1}{N_A^2}\Bigg(\sum_{j=1}^{N_A}e^{-i\tau\lambda_j}\Bigg)\Bigg(\sum_{k=1}^{N_A}e^{i\tau\lambda_k}\Bigg)
\nn\\
&=&\frac{1}{N_A^2}\bigg(\mathrm{Tr}_A\rho_A^{i\tau}\bigg)\bigg(\mathrm{Tr}_A\rho_A^{-i\tau}\bigg)
\nn\\
&=&\frac{1}{N_A^2}\bigg|\mathrm{Tr}_A\rho_A^{i\tau}\bigg|^2
\nn\\
&=&\frac{1}{N_A^2}\bigg|e^{(1-i\tau)S_A(i\tau)}\bigg|^2
\nn\\
&=&\frac{\bigg|e^{(1-i\tau)S_A(i\tau)}\bigg|^2}{\bigg|e^{S_A(0)}\bigg|^2}
\nn\\
&=&1,
\eea
in which the Rényi entropy of a subregion $A$, $S_{A}(n)$, is defined by
\bea
S_{A}(n)\equiv \frac{\ln\mathrm{Tr}_A\rho_A^n}{1-n}.
\eea
Note that despite the $n$-sheet method only working for integer $n$, we assume that such analytical continuation exists even for non-integer $n$. 
Here we consider the central charge term, which is the dominant term in the Rényi entropy \cite{Calabrese:2004eu}
\bea
S_A(n)=\bigg(1+\frac{1}{n}\bigg)A(L),
\eea
where $A(L)$ is a real-valued function of an interval with length $L$. 
The form of $A(L)$ does not change our result. 
Note that there is a non-universal coefficient $c_n$, like 
\bea
\mathrm{Tr}\rho_A^n\sim c_n\cdot f_n(L),
\eea
with $c_1$ being unity appearing in the R\'enyi entropy. 
Although the coefficient also depends on the index $n$, we can ignore it in the limit
\bea
\frac{L}{\epsilon}\rightarrow\infty,
\eea
where $\epsilon$ is a regularization parameter. 
Previously, we showed that the spectral form factor is one for Bell's state. 
A single interval in CFT$_2$ does not correspond to Bell's state, where its Rényi entropies depend on index $n$.
Let us also apply the calculation to the supersymmetric Rényi entropy in the $\mathcal{N}=4$ super Yang-Mills theory \cite{Huang:2014pda}. 
The supersymmetric Rényi entropy is \cite{Huang:2014pda}
\bea
&&\frac{S_A(n)}{S_A(1)}
\nn\\
&=&
\frac{(a_1^2+a_1a_2-3a_1)(n-1)\big(3n-a_2(n-1)\big)}{27n^2}
\nn\\
&&+\frac{3n\big(a_2(a_2-3)(n-1)+9n\big)}{27n^2},
\eea
where $a_1$ and $a_2$ are some real constants. 
Note that the result is for the twisted reduced density matrix.
We define the spectral form factor by replacing the reduced density matrix with the twisted one. 
This result also gives one. 
The $\mathcal{N}=4$ super Yang-Mills theory and CFT$_2$ are widely understood. 
These theories are calculable in the context of AdS/CFT correspondence \cite{Maldacena:1997re}. 
Hence this non-trivially constrains the entanglement spectrum from the spectral form factor. 
The situation could be very different for a continuum state versus a discrete one. 
It is then crucial to examine their differences.
We will now give a discussion about the continuum case. 
\\

\noindent
For a continuum spectrum, the reduced density matrix of region $A$ is 
\bea
\rho_A=\int dk\ O(k)|k\rangle\langle k|.
\eea
We label different states by $k$ (it is not the eigenvalue of a state). 
For the continuous spectrum, the value of $O(k)$ can be larger than one with the requirement: 
\bea
\mathrm{Tr}_A\rho_A=1=\int dk\ O(k).
\eea
The discretized form should be clear
\bea
1=\sum_k O(k)\delta k 
\eea
Therefore, we find 
\bea
O(k)\sim\frac{1}{\delta k},
\eea
which is different from the discretized form of a discrete spectrum
\bea
1=\sum_k O(k),
\eea
where the values of $O(k)$ are smaller than or equal to one.
Note that a reduced density matrix is positive definite. 
The eigenvalues cannot be larger than one for a discretized spectrum.
It is not necessarily the case for the continuum spectrum. 
The result shows the difference between a continuum spectrum and a discrete spectrum.
We now use the continuum spectrum to investigate what the entanglement spectrum gives in CFT$_2$.
\\

\noindent
The spectral form factor with a modular Hamiltonian for a continuum spectrum is
\bea
g_A(\tau)\sim \int dkd\tilde{k}\ 
e^{-i\tau\big(\lambda(k)-\lambda(\tilde{k})\big)},
\eea
which is up to a normalization constant. 
When the spectral form factor with a modular Hamiltonian is one, the spectrum constrained as the following
\bea
\lambda(k)=\lambda(\tilde{k}),
\eea 
but the result is not implying that the Rényi entropy does not depend on the Rényi index. 
In QFT, the relation between the modular Hamiltonian and reduced density matrix also needs to be deformed, similar to what we discussed,
\bea
H_{\mathrm{mod}}=-\ln(\rho_A\cdot \delta k).
\eea  
Hence this implies that the eigenvalues of the reduced density matrix satisfy the following relation
\bea
\mathcal{O}(k)\delta k=\mathcal{O}(\tilde{k})\delta\tilde{k}.
\eea
The eigenvalue $\mathcal{O}$ is relevant to the regularization. 
When we choose
\bea
\delta k=\delta\tilde{k},
\eea
the continuum case is equivalent to the discrete case. 
In general, they give different results. 
When one does a calculation in CFTs, conformal mapping is necessary.
This mapping should map a uniform discretization to a non-uniform lattice. 
The lattice spacing is the same as a regularization parameter. 
Hence it is why the $n$-dependence appears in the Rényi entropy. 
When the sum in a reduced density matrix is discrete (discrete spectrum), the fact that
\bea
g_A(\tau)=1
\eea
implies that the Rényi entropy does not depend on the Rényi index. 
It is in contrast with a continuous sum (continuous spectrum).
Usually, QFT has an infinite degree of freedom. 
If one can map QFT to the lattice model (like Toric Code), the result of the discrete spectrum can apply to QFT.
However, the result is non-trivial. 
Our study indicates the difference between the continuum and discrete spectrum and thus constraining the entanglement spectrum.
Our technique does not restrict to CFT$_2$. 
One can study a more generic QFT like $\mathcal{N}=4$ super Yang-Mills theory. 
 
\section{Discussion and Conclusion}
\label{sec:4} 
\noindent
We replaced the Hamiltonian with a modular Hamiltonian. 
It is advantageous to study spectral form factor from Quantum Entanglement.
The maximum violation of Bell's inequality shows a positive correlation to entanglement entropy in a 2-qubit state \cite{Verstraete:2001}.
Therefore, we first discuss the Gaussian random two-qubit model.
We first gave an analytical solution for the occurring time of the first dip. 
When entanglement entropy increases, the occurring time of the first dip increases. 
This estimation provides an interpretation of quantum chaos behavior from Quantum Entanglement. 
A consistent study can also be obtained from the 3-qubit case. 
We observed that the late-time dynamics is independent of the choice of an initial state. 
Such independence is one condition of classical chaos. 
We compared the spectrum to the random matrix theory \cite{Dyson:1962es} by calculating the level spacing distribution function. 
The result indicated GUE. 
Therefore, we conclude that randomness generates the statistical ensemble. 
Because Quantum Entanglement does not have a classical analogy, the chaotic behavior should be due to the complexity.   
In the end, we calculated the spectral form factor (for a single interval in CFT$_2$ and a spherical entangling surface in $\mathcal{N}=4$ super Yang-Mills theory). 
Both cases give 1 as a consequence. 
The results strongly constrain the entanglement spectrum. 
We also showed the difference between the continuum and the discrete spectrum. 
Hence one cannot naively extend the result from the random qubit models to QFTs. 
\\
 
\noindent
Quantum Chaos does not have a clear definition. 
We hope to search for solutions from a subregion. 
Thus, a path worth exploring is to probe the relation from an integrable system. 
For example, one can integrate over a subregion to obtain a non-local theory. 
Studying classical dynamics is possible to find explicit evidence of the connection. 
\\

\noindent
We considered a single interval in 2d CFT and a spherical entangling surface in $\mathcal{N}$=4 super Yang-Mills theory. 
The spectral form factor with the modular Hamiltonian is 1. 
It gives an example of constraining the entanglement spectrum. 
In the context of AdS/CFT correspondence \cite{Maldacena:1997re}, it should be particularly non-trivial because one can look for any universal constraint from the entanglement spectrum. 
The CFT$_2$ result implies the difference between the discrete and continuum cases.
Hence the discrete holographic tensor network might not be a complete statement about the AdS/CFT correspondence. 

\section*{Acknowledgments}
\noindent
We thank Pak Hang Chris Lau for his participation in the qubit model. 
We also thank Su-Kuan Chu, Wu-Zhong Guo, Xing Huang, and Rong-Xin Miao for their discussions. 
Chen-Te Ma would like to thank Nan-Peng Ma for his encouragement.
\\

\noindent 
Chen-Te Ma acknowledges the YST Program of the APCTP; 
Post-Doctoral International Exchange Program; 
China Postdoctoral Science Foundation, Postdoctoral General Funding: Second Class (Grant No. 2019M652926); 
Foreign Young Talents Program (Grant No. QN20200230017).
Chih-Hung Wu acknowledges the National Science Foundation under Grant No. 1820908; 
the Ministry of Education, Taiwan.
 \\
 
 \noindent
 We thank the Kadanoff Center for Theoretical Physics for the hospitality of this work.


\end{document}